\documentclass[twocolumn,showpacs,superscriptaddress,preprintnumbers,amsmath,amssymb]{revtex4-2}
\usepackage{graphicx}
\usepackage{dcolumn}
\usepackage{appendix}
\usepackage{bm}

\usepackage[usenames,dvipsnames]{xcolor}
\usepackage{subfigure}
\usepackage{bbm}
\usepackage{enumerate}
\usepackage[
  bookmarks=true,
  colorlinks,
  linkcolor=blue,
  urlcolor=blue,
  citecolor=blue,
  plainpages=false,
  pdfpagelabels,
  final,
  breaklinks=true
]{hyperref}

\usepackage{titlesec}
\usepackage{array}
\usepackage{dsfont}
\usepackage{physics}
\usepackage{mathpazo}

\begin{document}

\title{Information-theoretic perspective on energy conservation in high harmonic generation}
\author{Philipp Stammer}
\email{philipp.stammer@icfo.eu}
\affiliation{ICFO -- Institut de Ciencies Fotoniques, The Barcelona Institute of Science and Technology, 08860 Castelldefels (Barcelona), Spain}
\affiliation{Atominstitut, Technische Universität Wien, Stadionallee 2, 1020 Vienna, Austria}

\date{\today}

\begin{abstract}

The use of energy conservation arguments is ubiquitous in understanding the process of high harmonic generation, yet a complete quantum optical description of exact photon number exchange remained elusive. Here, we solve this gap in description by introducing the energy conserving subspace in high harmonic generation in which many photons of the driving field are absorbed to generate a single photon of higher energy. The presented solution to energy conservation in quantum optical high harmonic generation naturally results in highly entangled states of light with non-classical properties in their marginals and photon statistics. This new technique can be seen as an information-theoretic approach to the problem of photon exchange between field modes, providing a new kind of selection rule imposed on the quantum optical state by the structure of the Hilbert space. In addition to providing the quantum state satisfying exact energy conservation, it allows to explain recent experimental results for quantum state engineering of optical cat states.

\end{abstract}

\maketitle

The notion of energy conservation is ubiquitous in the description of the process of high harmonic generation (HHG), in which the photons of an infrared (IR) driving field are upconverted to harmonic photons of higher energy~\cite{ferray1988multiple, l1993high}.
In all approaches to the explanation of the measured HHG spectrum arguments based on energy conservation are used.
This includes classical descriptions, in which the energy of the laser driven electron is considered~\cite{corkum1993plasma}, and is further resumed in the semi-classical approach using classical light fields~\cite{lewenstein1994theory}. 
Although the concept of the photon is ambiguous without quantizing the radiation field, the parametric nature of the HHG process leads to explanations using arguments of the number of absorbed and emitted photons. Due to conservation of energy and parity only odd harmonics of the fundamental frequency are observed~\cite{perry1993high}, which is further extended to the appearance of even harmonics in two-color driving fields~\cite{pisanty2014spin}.
Despite all these arguments are based on energy conservation and use a photon picture, there is, however, no consistent quantum optical description in terms of quantum states and photon numbers which take into account energy conservation of the process. 

Nevertheless, there is the recently emerging field of quantum optical approaches to HHG which are rapidly developing~\cite{gorlach2020quantum, lewenstein2021generation, stammer2023quantum, cruz2024quantum, stammer2025theory}, but admittedly follow a different aim with the main focus in generating non-classical states of light~\cite{cruz2024quantum, stammer2025colloquium}.
Initial studies provided insights about the quantum state of the depleted IR driving field and of the generated harmonic radiation~\cite{lewenstein2021generation}, resulting in classical product coherent states for simple atomic gas targets and moderate driving laser intensity.
However, there are many different path predicting how to generate non-classical field states. Instead of using simple atomic targets it was shown that quantum correlated materials can lead to non-trivial Wigner functions~\cite{pizzi2023light, stammer2025high}, and non-classical properties in the harmonic field modes~\cite{lange2024electron}. It was also shown that depletion of the electron ground state due to high laser intensities leads to entanglement and squeezing in the optical field modes due to electron correlations~\cite{stammer2024entanglement}, or the presence of resonant media allows to generate massive entangled state~\cite{yi2024generation}.
Alternatively, using a squeezed light driving field~\cite{gorlach2023high, rasputnyi2024high}, allows to shape electron trajectories in the continuum~\cite{even2023photon}, and can induce squeezing in the generated harmonics~\cite{tzur2024generation, rivera2025attosecond}.
Furthermore, experiments have revealed entanglement between the optical field modes in HHG from semiconductors~\cite{theidel2024evidence, theidel2024observation}.
In addition, there are experimental approaches using conditioning measurements in HHG~\cite{tsatrafyllis2017high}, which allow to generate non-classical field states in the IR driving field~\cite{lewenstein2021generation, rivera2022strong}. These conditioning schemes rely on post-selection on the HHG process~\cite{stammer2022high, stammer2022theory}, such that these energy conserving events are isolated~\cite{rivera2024quantum}. By tomographic reconstruction of the Wigner function of the field it was shown that the state resemble an optical cat state of high average photon number~\cite{lewenstein2021generation, rivera2024quantum}. However, a precise description of the experiment without further assumptions remained elusive.
While all aforementioned work provide tremendous progress for the field in describing the quantum state after HHG, they all fail to consider energy conservation in a satisfactory and precise way.

Now, in this work we achieve both, we provide a precise description of exact energy conservation in the quantum optical context, and show how this naturally leads to entangled states of light with non-classical properties in their marginals and photon statistics.
This is achieved by introducing the new technique of the \emph{energy conserving subspace} for the process of HHG in terms of photon number states. The underlying physical picture formally takes into account the exact exchange of photon numbers between the different field modes, and therefore automatically satisfies energy conservation.
This can be seen as an information-theoretic approach to provide a new kind of selection rule imposed on the quantum state satisfying exact energy conservation.
On top of solving the energy conservation description in HHG from a quantum optical perspective, we get the explanation of the recent experimental results of conditioning experiments in HHG for free~\cite{lewenstein2021generation}.
In the following, we will first set the stage with the formal description of the quantum state after HHG, and discuss the difficulty in defining energy conservation in the microscopic picture. We then introduce the energy conserving subspace for HHG to solve this problem, and discuss the consequences for generating non-classical states of light and the structure of the quantum state.

Starting with the output state from the HHG process, given as a product of coherent states induced by a classical charge current of the laser driven electron~\cite{lewenstein2021generation, stammer2023quantum}, we have 
\begin{align}
\label{eq:state_wavepicture}
    \ket{\psi} = \ket{\alpha + \delta \alpha} \otimes \ket{\chi_2} \otimes ... \otimes \ket{\chi_{q_c}},
\end{align}
where the amplitude of the initial driving field mode $\alpha$ is shifted by $\delta \alpha$ and the harmonic field modes $q$ are in coherent states with amplitude $\chi_q$ until the harmonic cut-off $q_c$. The coherent states are induced by the classical charge current of the laser driven electron, and the state of Eq.~\eqref{eq:state_wavepicture} is obtained from the solution of the Schrödinger equation under the assumption of moderate laser intensity in a coherent state $\ket{\alpha}$, and negligible depletion of the ground state leading to vanishing dipole correlations~\cite{lewenstein2021generation, stammer2023quantum, cruz2024quantum, stammer2024entanglement}. 
Giving that the output state is in a product of coherent states, there are no correlations between the different modes. Therefore, fluctuations in the photon number of the modes are independent of each other and measurements performed on them.
Furthermore, the solution in Eq.~\eqref{eq:state_wavepicture} is not consistent with energy conservation arguments. This is because the depletion in the driving field is formally independent of the harmonic amplitudes.
In this context, it is important to note that from an energy conservation perspective the depletion of the driving field amplitude $\delta \alpha$ should be a function of the generated harmonic amplitudes $\delta \alpha( \{ \chi_q \} )$, where $\{ \chi_q \} = \{ \chi_2, \chi_3, ... , \chi_{q_c} \}$.

However, looking at the description of HHG via Eq.~\eqref{eq:state_wavepicture}, this dependence is neither explicitly given nor does the state include any correlations between the different field modes. 
The photon numbers, and their fluctuations, of the harmonic field modes are independent of the reduced photon number in the driving field. 
This is in contrast to the intuition that the number of photons generated in the harmonic field modes are correlated to the missing photons in the driving IR field. Even in the semi-classical picture of HHG, in which a classical field drives the process, and thus not admitting the notion of a photon, such an energy conservation picture is frequently used~\cite{lewenstein1994theory}. 
For instance, the appearance of only odd harmonics ($q \bmod 2 = 0 $) is well explained by energy and parity conservation during the total exchange of photons between the modes, i.e. for the emission of a single photon of mode $q$ only an odd number of IR photons can be absorbed such that the total parity does not change~\cite{bertrand2011ultrahigh}. Similar arguments based on photon exchange and conservation of angular momentum and energy conservation is used to explain the structure of the HHG spectrum when driven by two-color fields~\cite{pisanty2014spin, fleischer2014spin}. 
However, so far there is no consistent quantum optical description in terms of the quantum state of the field which takes into account energy conservation on the exact photon number level. 
In the following, we solve this problem and rigorously show how energy conservation for HHG can be included in the quantum optical setting. 
This is done by introducing the concept of the \emph{energy conserving subspace} for the up-conversion process of HHG in which IR photons of energy $\omega$ are converted into harmonic photons of energy $\omega_q = q \omega$. 
The correlated photon absorption and emission process has the natural consequence that the quantum state of the light field is entangled, and shows non-classical photon statistics.

Before we start introducing the energy conserving subspace in HHG, we provide some preliminary understanding about the decomposition of the total Hilbert space $\mathcal{H}$. 
Therefore, assume we have $n$ systems of dimension $d$ each, such that the total Hilbert space is given by $\mathcal{H} = \mathcal{H}^{(1)} \otimes ... \otimes \mathcal{H}^{(n)} = (\mathds{C}^d)^{\otimes n}$.
The identity on the total Hilbert space is simply $\mathds{1}_{d^n} = \mathds{1}_d \otimes ... \otimes \mathds{1}_d$, and we shall decompose the total Hilbert space into different subspaces. 
First, let us do an illustrative example. Therefore, we shall consider the case of $n=2$ systems of dimension $d=3$, such that the identity reads
\begin{align}
    \mathds{1}_9  = \sum_{n=0}^2 \dyad{n} \otimes \sum_{m=0}^2 \dyad{m},
\end{align}
where we have used the resolution of the identity in terms of particle or photon number states $\dyad{n}$. 
By inspection of the different terms, we can decompose the identity into subspaces of different photon number.  
For instance, the subspace of $N=2$ photons is spanned by $\operatorname{span}(\{ \ket{11} , \ket{20} , \ket{02} \})$, and all projectors $\mathcal{P}^{(N)}$ onto subspaces of different photon numbers $N$ resolve the identity $\mathds{1} = \sum_{N=0}^{N_{max}}  \mathcal{P}^{(N)}$, where the maximal photon number is given by $N_{max} = n (d-1)$.

However, here, we are not interested in the photon number conserving subspaces, but rather in the different subspaces which conserve the total energy $\mathfrak{E}$. We therefore define with $\Pi^{(N \omega)}$ the projector on the subspace of total energy $N \omega$, such that 
\begin{align}
    \mathfrak{E} \left[ \Pi^{(N \omega)} \right] = \frac{ \Tr[ \sum_q \omega_q a_q^\dagger a_q \Pi^{(N \omega)}  ] }{\operatorname{dim}[\Pi^{(N \omega)}]}  =  N \omega.
\end{align}

This decomposes the Hilbert space into subspaces of different total energy, such that within each subspace the total energy is preserved. Note that if all field modes have the same energy this would correspond to a resolution into subspaces of different photon number with each subspace being a photon number conserving subspace (and therefore being the energy conserving subspace if $\omega_q = \omega_{q+1} \, \forall q$).
Thus, the photon number preserving subspace is also the energy conserving subspace for photons of the same energy in each subsystem, i.e. field mode.
However, non-linear processes and specifically HHG are different since each field mode has a different energy of $\omega_q = q \omega$. We shall therefore look at the resolution of the identity into subspaces of different energy by taking into account the energy per photon in each mode.
To get an idea of this energy conserving subspace we shall continue to consider the case of a bipartite system of dimension $d=3$, and assume second harmonic generation (SHG) such that the first mode has energy $\omega$ per photon and the second mode has a photon energy of $2 \omega$. We can thus decompose the identity into different energy conserving subspaces 
\begin{align}
    \mathds{1}_{9} = \sum_{N=0}^6  \Pi^{(N\omega)},
\end{align}
where each projector $\Pi^{(N \omega)}$ corresponds to the subspace of equal energy. To be colorful we provide the projectors for different subspaces, such as $\Pi^{(2\omega)} = \dyad{20} + \dyad{01}$ or $\Pi^{(4\omega)} = \dyad{21} + \dyad{02}$, 
where the energy of the respective subspaces is given by $\mathfrak{E}[\Pi^{(N \omega)}] = N \omega$, and for the highest energy subspace we have $\mathfrak{E}[\Pi^{(6 \omega)}] = 2 \omega + 2 (2 \omega) = 6 \omega$.
To illustrate the field state in a given energy conserving subspace, we shall briefly consider the projection onto the subspace of energy $2\omega$, which for an arbitrary state $\ket{\lambda}$ is given by $\Pi^{(2\omega)} \ket{\lambda} = \lambda_{20} \ket{20} + \lambda_{01} \ket{01}$,
with $\lambda_{20} = \bra{20}\ket{\lambda}$ and $\lambda_{01} = \bra{01}\ket{\lambda}$.
This represents an entangled state between the field modes conditioned on the subspace of fixed energy. Here, the energy is either distributed on the fundamental mode including 2 photons of energy $\omega$ and no photon in the second harmonic mode ($\ket{20}$), or after generation of a single second harmonic photon of energy $2\omega$ with the fundamental mode in the vacuum after absorption of two photons ($\ket{01}$).

However, the process of HHG is more complex than that of SHG by means of the number of field modes involved, and how the photon energy in the harmonic modes is distributed. In HHG we have the excitation of many field modes of higher orders of the fundamental photon energy until a cutoff harmonic $q_c$, where the harmonics are odd multiple of the driving field frequency.

For the sake of the argument we shall consider the case of a single harmonic field mode $q$ with energy $\omega_q = q \omega$, such that the energy conserving subspace of energy $N \omega$ is defined by the projector
\begin{align}
    \Pi^{(N \omega)} = & \dyad{N, 0} + \dyad{N-q, 1} + ... \\
    & ... + \dyad{N \bmod q , \lfloor N/q \rfloor}. \nonumber 
\end{align}

\begin{figure}
    \centering
	\includegraphics[width=0.99\columnwidth]{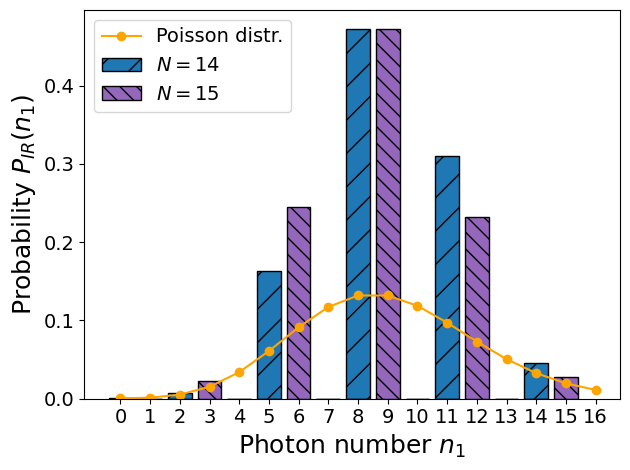}
    \caption{\textbf{Photon number distribution of the energy conserved state.} $P_{IR}(n_1)$ of the fundamental field when constrained to energy conservation on the subspace $\Pi^{(N\omega)}$, for the different energy subspaces $N \in [14, 15]$. The solid line corresponds to the Poisson distribution of the unconstrained state $\ket{\psi}$ (the coherent state amplitudes are given by $\alpha + \delta \alpha = 3.0$, and $\chi_q = 1.5$).}
    \label{fig:photon_number}
\end{figure}

We can now project the state from Eq. \eqref{eq:state_wavepicture} onto the subspace of energy $N \omega$, and we obtain 
\begin{align}
\label{eq:total_EspaceN}
    \Pi^{(N \omega)} \ket{\psi} = \sum_{n_q=0}^{\lfloor N/q \rfloor} c_{N-qn_q, n_q} \ket{N-qn_q , n_q},
\end{align}
with the coefficients of the subspace expansion given by $c_{i,j} = \bra{i,j} \ket{\psi} = \bra{i} \ket{\alpha + \delta \alpha} \bra{j} \ket{\chi_q}$.
This constitutes a highly entangled state between the two field modes, and provides the energy conserving state in HHG.
The first main insight obtained from this, is the fact that the quantum state after HHG is entangled when energy conservation is taken into account seriously, in contrast to the widely considered product state~\cite{lewenstein2021generation, gorlach2023high}.
Note, however, that we have so far assumed that the state is confined to the energy subspace $\Pi^{(N\omega)}$, although the dynamics of HHG is driven by a coherent state $\ket{\alpha}$ without a well defined photon number. We shall return to this \emph{absence of knowledge} about the exact energy conserving subspace further below. 
For now, and before we continue to investigate the structure of the quantum optical state under the constraint of energy conservation, we first gain some physical intuition on the implications of strict energy conservation. 
For this, given the normalized energy conserving state $\ket{\Psi_N} \equiv \Pi^{(N \omega)} \ket{\psi} / \sqrt{A}$, we can look at the photon number distribution of the fundamental IR mode
\begin{align}
\label{eq:Pn_distribution}
    P_{IR} (n_1) &= \Tr[(\dyad{n_1} \otimes \mathds{1}) \dyad{\Psi_N}], 
\end{align}
which is a conditional probability for observing $n_1$ photons in the fundamental mode given that the harmonic mode photon number satisfies $n_q = (N-n_1)/q \in \mathbb{N}$.
In Fig.~\ref{fig:photon_number} we show the photon number distribution of Eq.~\eqref{eq:Pn_distribution} for different energy subspaces $N \in [14, 15]$ and $q=3$, in comparison to the Poisson distribution of the product coherent state $\ket{\psi}$. 
We can directly see that individual energy subspaces have a non-vanishing support on different photon numbers. This is due to the energy conservation constraint placed on the photon number of the driving field, such that the only contributing photon numbers obey $n_1 = N -q n_q \in \mathbb{N}$. While an individual energy subspace is far from the Poisson distribution of the coherent state (solid line) due to the missing occupations of specific photon numbers, the joint distribution of all energy subspaces comes closer to the Poisson case (note that a joint distribution needs to be re-normalized).
For completeness, we note that the distribution for the harmonic photons is trivial since all photon numbers $n_q$ are allowed (up to the maximum number of photons that can be generated $n_q^{max} = \lfloor N/q \rfloor$).

Having gained some physical intuition of the energy constraint, we now want to get an idea on the implications for the quantum optical state within the energy conserving subspace. Therefore, we analyze the driving laser mode in more detail, obtained when tracing Eq.~\eqref{eq:total_EspaceN} over the harmonic mode 
\begin{align}
\label{eq:fundamental_EspaceN}
    \rho = \frac{1}{P^{(N\omega)}} \sum_{n_q=0}^{\lfloor N/q \rfloor} \abs{c_{N-qn_q, n_q}}^2 \dyad{N-qn_q },
\end{align}
with the probability of being in the particular subspace $P^{(N \omega)} = \operatorname{Tr} [\Pi^{(N\omega)} \dyad{\psi}]$ for normalization. Note that the state is a statistical mixture of photon number states, and therefore diagonal in the photon number basis~\cite{stammer2024absence}.

\begin{figure}
    \centering
	\includegraphics[width=0.49\columnwidth]{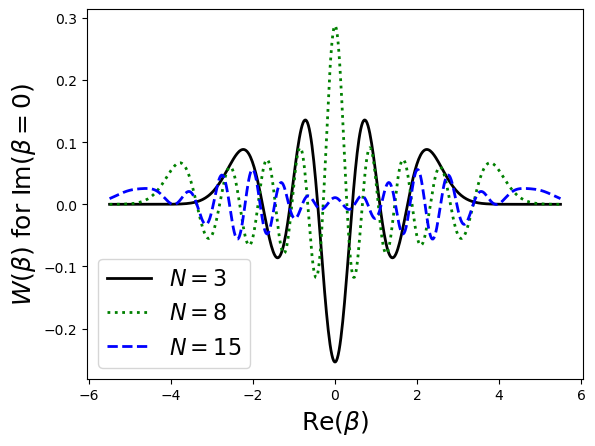}
	\includegraphics[width=0.49\columnwidth]{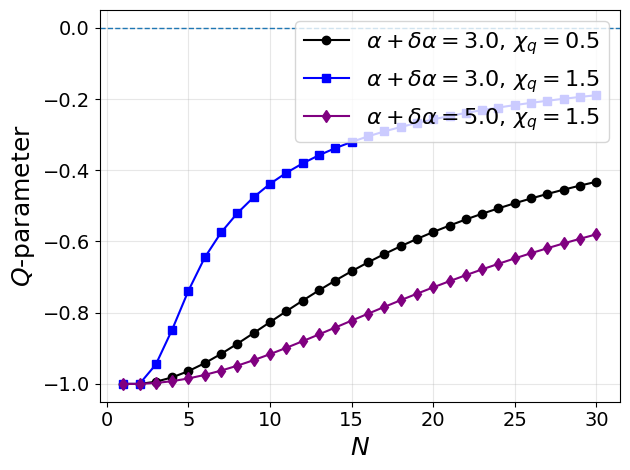}
    \caption{\textbf{Non-classical states in HHG.} Left: Marginal of the Wigner functions $W(\beta)$ for different energy conserving subspaces $\Pi^{(N \omega)}$ of Eq.~\eqref{eq:fundamental_EspaceN}, with total energy $N \omega = 3 \omega$ (solid line), $8 \omega$ (dotted) and $15 \omega$ (dashed line). For the respective subspaces the coherent state amplitudes of the state in Eq.~\eqref{eq:state_wavepicture} are $ \{\alpha = 1.2, \delta \alpha = -0.3, \chi_q = 0.1 \}_3 $,  $\{4.2, -1.3,0.3 \}_8 $, and  $\{ 5.2,  -2.3, 0.8 \}_{15} $. Right: Mandel $Q$-parameter of the photon number distribution from Fig.~\ref{fig:photon_number} as a function of the energy subspace $N\omega$, indicating clear sub-Poissonian statistics. }
    \label{fig:wigner}
\end{figure}

We now illustrate the state by showing the trace of the Wigner function $W(\beta)$ along $\operatorname{Im}(\beta) = 0$ in Fig.~\ref{fig:wigner}~(left), given by a weighted sum of the individual photon number state Wigner functions $W_{\dyad{n}}(\beta) = \frac{(-1)^n}{\pi} e^{- \abs{\beta}^2} L_n(2\abs{\beta}^2)$, where $L_n(x)$ is the Laguerre polynomial of $n$-th order.
We show the Wigner function $W(\beta)$ of the reduced state in Eq.~\eqref{eq:fundamental_EspaceN} for different energy conserving subspaces $N = \{ 3, 8, 15 \}$. We can clearly see the oscillatory behavior of the Wigner function originating from the high photon number states, and that the structure of the state strongly depends on the specific choice of the subspace. However, most importantly, we observe that all states display clear negativities in their Wigner function representation, highlighting that we can witness non-classicality in the marginals of the entangled state in Eq.~\eqref{eq:total_EspaceN}.
Furthermore, to get additional insights into the non-classicality of the driving field mode, we show in Fig.~\ref{fig:wigner}~(right) the Mandel $Q$-parameter of the photon number distribution for varying energy subspace $N\omega$ (see Fig.~\ref{fig:photon_number}). The $Q$-parameter is a measure of the deviation of the photon number distribution from the Poissoian ($Q=0$) case, and is defined as the normalized difference between the variance and the mean~\cite{mandel1979sub}. For $Q< 0$ the distribution is sub-Poissonian, indicating non-classical statistics. We can clearly see that for all energy subspaces $N\omega$, the photon statistics is non-classical due to the negative $Q$-parameter. For increasing $N$ the distribution gets closer to be Poissonian since more allowed photon numbers contribute to the distribution.
From the results shown in Fig.~\ref{fig:wigner}, we have learned that strict energy conservation leads to non-classical states of the light field. This can be seen as a new kind of selection rule imposed on the quantum optical state due to energy conservation.

Having established the non-classicality of the energy conserving state, we shall now discuss a crucial aspect about the energy conserving subspace as defined in Eq.~\eqref{eq:total_EspaceN}, which is the important assumption of a well defined energy $N\omega$ on the photon number level.
However, considering that the driving field in HHG is given by a coherent state $\ket{\alpha}$, implies that the photon number is not fixed but follows a Poisson distribution. 
Hence, the energy can not be conserved on the exact photon number level since the initial photon number of the coherent state is not well defined. This is because the coherent state is not an energy eigenstate of the field Hamiltonian.
This implies that the initial energy conserving subspace is not unambiguously defined, and due to the indistinguishability of which particular energy conserving interaction takes place, we project the state in Eq.~\eqref{eq:state_wavepicture} on several energy conserving spaces $N\omega$ with $N \in [ N_0 - \Delta N, N_0 + \Delta N]$, such that we have 
\begin{align}
\label{eq:ECS_summed}
    \sum_{N} \Pi^{(N \omega)} \ket{\psi} = \sum_{N} \sum_{n_q=0}^{\lfloor N/q \rfloor} c_{N-qn_q, n_q} \ket{N-qn_q , n_q}.
\end{align}

From the physical intuition of the photon number distribution of the state $\Pi^{(N\omega)}\ket{\psi}$ for different energy constraints from Fig.~\ref{fig:photon_number}, we can already anticipate the influence of the summation over $N$. Given that the neighboring energy spaces lead to an occupation of different photon numbers, their summed contribution will consequently cover more photon numbers. 
For an increasing summation over energy subspaces $N$, we sample more states from the entire Fock space. Since different energy subspaces are mutually orthogonal, and provide a resolution of the identity, the overlap with the unconstrained state $\ket{\psi}$ will therefore increase as more energy subspaces are taken into account.

Before concluding on the importance of the energy conserving subspace in HHG, we briefly discuss the post-selection experiment introduced in Ref.~\cite{tsatrafyllis2017high}, where conditional measurements on HHG are performed. These experiments have later been extended to the quantum optical framework of HHG to experimentally reconstruct non-Gaussian states of the IR driving field as reported in Ref.~\cite{lewenstein2021generation, rivera2022strong}. These experiments are of particular interest since they allow to generate non-classical states of the IR driving field by means of negatitivies in their Wigner function representation. These states are associated with generalized optical cat states, given by superpositions of coherent states of the form $\ket{cat} = \ket{\alpha + \delta \alpha} - \bra{\alpha}\ket{\alpha + \delta \alpha} \ket{\alpha}$, and supposedly provide the brightest optical cat state to date~\cite{lamprou2023nonlinear}. 
However, existing descriptions either use phenomenological explanations~\cite{lewenstein2021generation}, or a numerical sampling approach of the experiment~\cite{rivera2024quantum}. Using the energy conserving subspace introduced in this work, allows to fill these gaps in the explanation by providing a complete description of the experiment. 
While a detailed derivation can be found in the Supplementary Material, we emphasize that the approach with the energy conserving subspace yields a fidelity of $F=0.998$ between the energy conserving state and the cat state reported in Ref.~\cite{lewenstein2021generation}.

In conclusion, we have introduced the energy conserving subspace in the process of HHG, which describes the quantum state of the total light field by means of the exact photon exchange between the modes.
It therefore provides a rigorous way of describing energy conservation based on exact photon numbers in HHG from a quantum optical standpoint, a so far elusive picture in the semi-classical description and all existing full quantum approaches.  
This photon picture reveals the immanent correlations between the field modes in the HHG process, due to the absorption and emission of photons, and highlights the entanglement between the modes leading to non-classical field statistics. 
Imposing energy conservation constraints on the final quantum optical state of the field, provides a new kind of selection rule for the number of absorbed photons of the driving field for generating a specific number of harmonic photons. Selection rules in HHG have been considered from a dynamical standpoint~\cite{lerner2024reflection}, while in this work the problem at hand is approached using information theoretic concepts by partitioning the Hilbert space. We expect that an extension of such an information theoretic perspective, by including additional degrees of freedom, such as spin angular momentum for bicircular fields, will lead to further interesting insights.

With the presentation of the energy conserving subspace in HHG, we made sense of previously diverging descriptions of the quantum state in HHG. In particular, we introduced a new kind of selection rule for the quantum state, and provided a straightforward explanation of the post-selection experiment introduced in~\cite{lewenstein2021generation}.
Since the method introduced in this work is generic to energy conserving processes, i.e. parametric processes such as second harmonic generation or four-wave mixing, the notion of the energy conserving subspace is of particular importance for a wide class of non-linear quantum optical processes~\cite{chang2014quantum}.

\begin{acknowledgments}

I thank the participants of the Extreme Quantum Optics Workshop in Barcelona in March 2024 for the discussions about HHG, which eventually inspired thinking leading to this work. 
P.S. acknowledges funding from the European Union’s Horizon 2020 research and innovation programe under the Marie Skłodowska-Curie grant agreement No 847517. 
ICFO group acknowledges support from: Ministerio de Ciencia y Innovation Agencia Estatal de Investigaciones (R$\&$D project CEX2019-000910-S, AEI/10.13039/501100011033, Plan National FIDEUA PID2019-106901GB-I00, FPI), Fundació Privada Cellex, Fundació Mir-Puig, and from Generalitat de Catalunya (AGAUR Grant No. 2017 SGR 1341, CERCA program), and MICIIN with funding from European Union NextGenerationEU(PRTR-C17.I1) and by Generalitat de Catalunya and EU Horizon 2020 FET-OPEN OPTOlogic (Grant No 899794) and ERC AdG NOQIA.

\end{acknowledgments}

\bibliography{literatur}{}

\clearpage
\onecolumngrid
\appendix

\section*{Supplementary Material}

\subsection{The energy conserving subspace for cat state conditioning experiments}

As highlighted in the main text, the technique of the energy conserving subspace allows for an explanation of the experiments performed in Refs.~\cite{lewenstein2021generation, rivera2022strong} without using phenomenological assumptions or postulates~\cite{rivera2024quantum}. This reinforces the predictive power of this new technique by bridging the gap of previous unfinished explanations, where the description of the experiments have thus far remained elusive.
Using the energy conserving subspace introduced in this work we are now able to provide a complete description of the experimental results without postulating specific measurement operators, and only assume energy conservation to hold in the process of HHG. 
The aforementioned cat state is of the form~\cite{lewenstein2021generation, rivera2022strong, stammer2022high, stammer2022theory}
\begin{align}
\label{eq:cat}
    \ket{cat} = \ket{\alpha + \delta \alpha} - \bra{\alpha}\ket{\alpha + \delta \alpha} \ket{\alpha},
\end{align}
and is given by a superposition of coherent states of unequal amplitude, which makes the state particularly robust against decoherence due to photon loss~\cite{stammer2024metrological}. 

These cat states are obtained when performing post-selection on the energy conserving events in HHG~\footnote{Details on the experimental implementation can be found in Ref.~\cite{stammer2023quantum}.}, which makes the energy conserving subspace a natural candidate for describing the experiment. 
Conditioning on the energy conserving events in the experiment is achieved when measuring the photon number in the harmonic field mode, and correlate the measured photons with the one measured in the driving field. 
If we project the entangled state in Eq.~(6) of the main manuscript on a measured photon number state of the harmonic $\ket{n_q}$, such that the fundamental field mode is in the photon number state, $\bra{n_q} \Pi^{(N\omega)} \ket{\psi} = \ket{N-qn_q}$. The driving field is now missing $qn_q$ IR photons within the energy conserving subspace $N\omega$. 
However, since the exact photon number of the initial driving field is unknown, we need to consider the state of several energy subspaces as defined in Eq.~(9) of the main manuscript. 
If we now project on a measured harmonic photon number, $\ket{\Phi(n_q)} = \bra{n_q} \sum_{N_i} \Pi^{(N_i \omega)} \ket{\psi}$, we have 
\begin{align}
\label{eq:IR_postselection}
    \ket{\Phi(n_q)}  = \sum_{N_i} c_{N_i-qn_q, n_q} \ket{N_i-qn_q},
\end{align}
which is a superposition of different photon number states. 
To explicitly compare these results with the aforementioned optical cat state in Eq.~\eqref{eq:cat}, we consider that the experiment performed in Ref.~\cite{lewenstein2021generation} is within the regime of small coherent state amplitudes. We therefore consider the energy conserving subspace of $N_0=10$ and variance $\Delta N = 1$, and measure in the $5$-th harmonic the maximum possible number of generated harmonic photons $n_{q=5} = 2$, such that the conditioned state of the driving field reads 
\begin{align}
\label{eq:fock_superposition_small}
    \ket{\Phi(n_{5}=2)} = c_{0,2} \ket{0} + c_{1,2} \ket{1}.
\end{align}

\begin{figure}
    \centering
	\includegraphics[width=0.75\columnwidth]{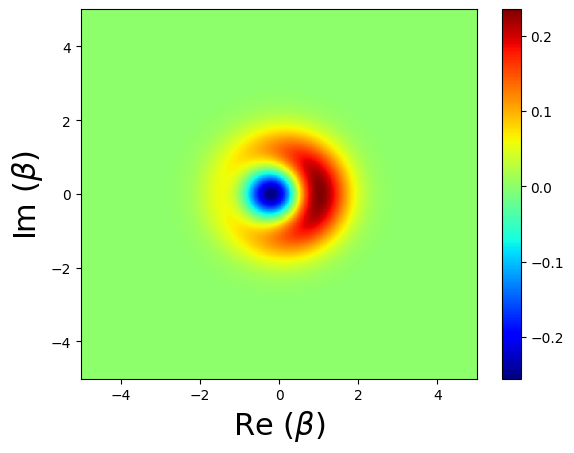}
    \caption{Wigner function $W(\beta)$~\cite{johansson2012qutip} of the normalized photon number superposition in Eq.~\eqref{eq:fock_superposition_small}, showing clear negativities and strong agreement with the previously reported experimental Wigner functions in Refs.~\cite{lewenstein2021generation, rivera2022strong}. The coherent state amplitudes are given by $\alpha = 2.5 $, $\delta \alpha = -0.1$ and $\chi_q = 0.1$. The state shown here has a fidelity of $F = 0.998$ with the cat state of Eq.~\eqref{eq:cat}.}
    \label{fig:wigner2d}
\end{figure}

We show the Wigner function of this photon number superposition in Fig.~\ref{fig:wigner2d}, exhibiting the same structure as the experimentally reconstructed Wigner function in Ref.~\cite{lewenstein2021generation, rivera2022strong}. 
To complete the analysis on the HHG conditioning experiment, and to assess the validity of the cat state description we use the fidelity $F(\ket{\psi}, \ket{\phi}) = \abs{\bra{\psi}\ket{\phi}}^2$ as a measure how close the pure cat state in Eq.~\eqref{eq:cat} is with the conditioned state in Eq.~\eqref{eq:IR_postselection}. 
We therefore have (note that we consider the normalized states such that the fidelity is upper bounded by $F=1$)
\begin{align}
\label{eq:fidelity}
    F & = \sum_{N_i} \abs{\bra{N_i-qn_q} \ket{\alpha + \delta \alpha}}^2 \frac{\abs{1- \xi^* \Delta P}^2}{(1- \abs{\xi}^2)}, 
\end{align}
where $\xi = \bra{\alpha} \ket{\alpha + \delta \alpha}$, and we have defined (see below for the detailed derivation)
\begin{align}
\label{eq:fidelity_delta}
    \Delta P = \frac{\sum_{N_i} \bra{\alpha} \ket{N_i - q n_q} \bra{N_i - q n_q} \ket{\alpha + \delta \alpha}  }{\sum_{N_i} \bra{\alpha +\delta \alpha} \ket{N_i - q n_q} \bra{N_i - q n_q} \ket{\alpha + \delta \alpha}}.
\end{align}

This allows to obtain upper and lower bounds for the fidelity in Eq.~\eqref{eq:fidelity}, given by 
\begin{align}
\label{eq:fidelity_bounds}
    \abs{1- \xi^* \Delta P}^2 \sum_{N_i} P_{N_i}(\alpha + \delta \alpha)  \leq F \leq \frac{\abs{1- \xi^* \Delta P}^2}{1- \abs{\xi}^2},
\end{align}
where $P_{N_i}(\alpha + \delta \alpha) = \abs{\bra{N_i-qn_q} \ket{\alpha + \delta \alpha}}^2$ is the probability if having $N_i-qn_q$ photons in the respective coherent state.
Inspecting the expression for the fidelity we can already get an idea on the requirements for a high fidelity. It is a careful balance between the different support of the two coherent states, $\ket{\alpha}$ and $\ket{\alpha + \delta \alpha}$, on the photon number states $\ket{N_i - qn_q}$, such that $\Delta P$ is minimized while $\sum_{N_i} P_{N_i}(\alpha + \delta \alpha)$ is maximized. 
To complete the analysis on the HHG conditioning experiment, we search for the optimal fidelity of the energy conserving state in Eq.~\eqref{eq:IR_postselection} and the cat state $\ket{cat} = \ket{\beta + \delta \beta} - \xi \ket{\beta}$. Specifically, we optimize over the coherent state amplitudes of the cat state $F(\beta^*, \delta \beta^*) = \max_{\beta, \, \delta \beta} \abs{\bra{cat (\beta, \delta \beta)}\ket{\Phi(n_q)}}^2$, where we consider the energy conserving subspace $N_0=10$ with $\Delta N = 1$, while keeping the same coherent state amplitudes for Eq.~\eqref{eq:IR_postselection} as shown in Fig.~\ref{fig:wigner2d}. The optimal value of the fidelity is given by $F=0.998$ for the cat state amplitudes of $\beta^* = -0.38$ and $\delta \beta^* = 0.70$, showing almost perfect overlap between the two states. 
With this, we have shown that the energy conserving subspace allows to explain the experiments performed in Refs.~\cite{lewenstein2021generation, rivera2022strong}, which have thus far resorted to phenomenological assumptions on the measurement operators. The energy conserving subspace $\Pi^{N\omega}$, in contrast, allows to formulate explicit measurement operations for these experiments. The conditioned state in Eq.~\eqref{eq:IR_postselection} is in fact obtained from Eq.~(1) of the main manuscript via the measurement operators~\cite{nielsen2010quantum, stammer2022theory}
\begin{align}
    M_{n_q}^{(N\omega)} \equiv \bra{n_q} \Pi^{(N \omega)} \ket{\chi_q}, 
\end{align}
which constitute a positive operator-valued measure (POVM), that satisfy the completeness relation
\begin{align}
    \sum_N \sum_{n_q=0}^\infty \left( M_{n_q}^{(N\omega)} \right)^\dagger M_{n_q}^{(N\omega)} = \mathds{1}.
\end{align}

\section{\label{app:fidelity}Derivation of fidelity}

Here we provide a more detailed derivation of the fidelity given in Eq.~\eqref{eq:fidelity} between the cat state $\ket{cat}$ in Eq.~\eqref{eq:cat} and the conditioned state in the energy conserving subspaces $\ket{\Phi(n_q)}$ in Eq.~\eqref{eq:IR_postselection}. The fidelity between the pure states is defined as 
\begin{align}
    F = \abs{\bra{cat} \ket{\Phi(n_q)}}^2. 
\end{align}

We shall first compute the overlap between the states 
\begin{align}
    \bra{cat} \ket{\Phi(n_q)} & = \frac{\bra{n_q}\ket{\chi_q}}{\abs{\bra{n_q} \ket{\chi_q}}} \frac{1}{\sqrt{1 - \abs{\xi}^2}} \frac{\sum_{N_i} \left[ \abs{\bra{\alpha + \delta \alpha} \ket{N_i - q n_q}}^2  - \xi^* \bra{\alpha} \ket{N_i - q n_q} \bra{N_i - q n_q} \ket{\alpha + \delta \alpha}\right]}{ \sqrt{ \sum_{N_i} \abs{ \bra{N_i - q n_q}\ket{\alpha + \delta \alpha} }^2 } } \\
    & = \frac{\bra{n_q}\ket{\chi_q}}{\abs{\bra{n_q} \ket{\chi_q}}} \frac{1}{\sqrt{1 - \abs{\xi}^2}} \left[ \sqrt{\sum_{N_i} \abs{\bra{\alpha + \delta \alpha} \ket{N_i - q n_q}}^2} - \frac{\xi^* \sum_{N_i} \bra{\alpha} \ket{N_i - q n_q} \bra{N_i -qn_q} \ket{\alpha + \delta \alpha} } { \sqrt{\sum_{N_i} \abs{ \bra{N_i - q n_q} \ket{\alpha + \delta \alpha} }^2} }  \right],
\end{align}
such that after taking the absolute square we get for the fidelity 
\begin{align}
    F & = \frac{\sum_{N_i} \abs{\bra{N_i-qn_q} \ket{\alpha + \delta \alpha}}^2}{(1- \abs{\xi}^2)} \left| 1 - \xi^* \frac{\sum_{N_i} \bra{\alpha} \ket{N_i - q n_q} \bra{N_i - q n_q} \ket{\alpha + \delta \alpha}  }{\sum_{N_i} \bra{\alpha +\delta \alpha} \ket{N_i - q n_q} \bra{N_i - q n_q} \ket{\alpha + \delta \alpha}} \right|^2 \\
    & = \frac{\sum_{N_i} \abs{\bra{N_i-qn_q} \ket{\alpha + \delta \alpha}}^2}{(1- \abs{\xi}^2)}  \abs{ 1 - \xi^* \Delta P }^2,
\end{align}
which coincides with Eq.~\eqref{eq:fidelity} when defining 
\begin{align}
    \Delta P = \frac{\sum_{N_i} \bra{\alpha} \ket{N_i - q n_q} \bra{N_i - q n_q} \ket{\alpha + \delta \alpha}  }{\sum_{N_i} \bra{\alpha +\delta \alpha} \ket{N_i - q n_q} \bra{N_i - q n_q} \ket{\alpha + \delta \alpha}}.
\end{align}

For the upper bound on the fidelity in Eq.~\eqref{eq:fidelity_bounds} we have used that 
\begin{align}
    \sum_{N_i} \abs{ \bra{N_i - q n_q} \ket{\alpha + \delta \alpha} }^2 & \leq  1,
\end{align}
such that 
\begin{align}
    F \leq \frac{\abs{ 1 - \xi^* \Delta P }^2}{(1- \abs{\xi}^2)}.
\end{align}

And for the lower bound, using $(1-\abs{\xi}^2) \leq 1$, we have 
\begin{align}
    F \geq \abs{ 1 - \xi^* \Delta P }^2 \sum_{N_i} P(N_i),
\end{align}
where we have defined $P(N_i) = \abs{\bra{N_i - qn_q} \ket{\alpha + \delta \alpha}}^2$.
We consequently have 
\begin{align}
    \abs{1- \xi^* \Delta P}^2 \sum_{N_i}P(N_i)  \leq F \leq \frac{\abs{1- \xi^* \Delta P}^2}{(1- \abs{\xi}^2)},
\end{align}
which is the result shown in Eq.~\eqref{eq:fidelity_bounds}.

\end{document}